\documentstyle{article}
\begin{document}


\begin{center}
{\large \bf QUANTUM MECHANICS IN GRASSMANN SPACE, SUPERSYMMETRY
AND GRAVITY}

\vspace{1.5cm}

       NORMA MANKO\v C BOR\v STNIK

\vspace{3mm}

{\it Department of Physics, University of Ljubljana, Jadranska 19, }\\
{\it J. Stefan Institute, Jamova 39,61 111 Ljubljana, Slovenia }\\

\vspace{12mm}

{\large ABSTRACT }

\end{center}

\vspace{3mm}

A particle which lives in a d-dimensional ordinary and a
d-dimensional Grassmann space  manifests itself in an ordinary
four-dimensional subspace as a spinor, a scalar or a vector with
charges. Operators of the Lorentz
transformations and translations in both spaces form the super-
Poincar\' e algebra. It is the super-Pauli-Ljubanski vector
which generates spinors.
Vielbeins and spin connections with the Lorentz index
larger than or equal to five may manifest in a four-dimensional
subspace as an electromagnetic, a weak and a colour field.

\begin{large}

\vspace{10mm}

1. INTRODUCTION

\vspace{5mm}

Not only we have  shown$^{1}$ that the supersymmetry can
consistently be formulated in the context of a Grassmann
space, but that this approach brings a new insight into the
concept of supersymmetry and physics, either in the flat space
or in the presence of gauge fields  entering into the
theory as vielbeins and spin connections.

We suppose that a particle lives in a d-dimensional ordinary and
a d-dimensional Grassmann space of anticommuting
coordinates. From the supersymmetric geodesics of the particle
it follows$ ^{1-2}$ that the action is invariant under the
Lorentz transformations in an ordinary and a Grassmann space
with the same transformation parameters in both spaces. The
canonical quantization of the dynamics of a particle in the
Grassmann space manifests in particle's internal degrees of
freedom ; a particle behaves like a scalar, a spinor or
a vector$^{1}$, with a weak or a colour charge$^{3}$, while the
momentum in the fifth dimension appears as an electromagnetic
charge. Generators of the Lorentz transformations in the
Grassmann space are differential operators in the Grassmann
space of an even Grassmann character and so are  the Dirac
$\gamma^{\mu}$ operators  which require $d\ge5$.

Vector space spanned over the Grassmann coordinate space has the
dimension $2^{d}$. Half of vectors have an even, half of vectors
have an odd Grassmann character. Canonical quatization of
fields quantizes the former to bosons, the later to fermions.

Generators of translations and the Lorentz transformations in
the ordinary and the Grassmann space  form the super-Poincar\' e
algebra$^{1}$. The super-Pauli-Ljubanski vector can be defined as
a generalization of the Pauli-Ljubanski vector with an odd
Grassmann character. It defines spinor charges.

Supervielbeins, depending on ordinary and Grassmann coordinates,
connect supervectors of a freely falling coordinate system to an
external coordinate system$^{1}$. A spin connection appears as a
superpartner of a vielbein. While a vielbein has an even
Grassmann character describing a spin 2 gravitational field, a
spin connection has an odd Grassmann character and describes a
fermionic part of a gravitational field. It is the dependence on
the Grassmann coordinates which determines the internal spin of
a gravitational field in the four dimensional subspace.

Vielbeins with a Lorentz index $a \ge 5$ manifest in the
four-dimensional subspace as gauge fields of Yang-Mills type,
the corresponding charges being defined by either the generators
of the Lorentz transformations in that part of the Grassmann
space which has the index higher than five (this is the case for
weak and colour charges) or by the momentum
in the fifth dimension of the ordinary space (which is the case
for the electromagnetic field).

A torsion and a curvature of the gravitational gauge field are
found by the Poisson brackets between components of covariant
momentum of a particle. Since veilbeins and spin connections
depend on ordinary and Grassmann coordinates, the derivatives
with respect to both types of coordinates appear in the Lagrange
density and correspondingly in equations of motion, defining
dynamics in both spaces.

\vspace{5mm}

2.  A PARTICLE IN A FREELY FALLING COORDINATE SYSTEM

\vspace{5mm}

We assume that a particle lives in a d-dimensional ordinary
$x^{a}$  and in a d-dimensional Grassmann space $\theta^{a}$ space
of anticommuting coordinates: $\lbrace x^{a},\theta^{a}\rbrace $
and that the action is invariant under the Lorentz
transformations with the same transformation parameters in
both spaces. We found two types of generators of the Lorentz
transformations corresponding to two different actions,
representations of one defining spinors, of the other defining
vectors.

For the action

$$ I=\int d\tau L(x^a, \theta^a,\dot{x} ^a, \dot{\theta} ^a)
\;,\;\;\dot{x} ^a= \frac{d}{d \tau} x^a \;,\;\; \dot{\theta}
^a= \frac{d}{d \tau} \theta^a, \eqno   (2.1) $$

in which $\tau$ is an ordinary time parameter, the generators of
the Lorentz transformations are

$$ M^{ a b} = L^{a b} + S^{a b} \;,\; L^{a b}= x^a p^b - x^b p^a
\;,\; S^{a b}= \theta^a p^{ \theta b} - \theta^b p^{ \theta a} ,
\eqno (2.2) $$

where
$$ p_a = \frac{\partial L}{ \partial \dot{x} ^a} \;,\;
p^\theta_a= \frac{\overrightarrow{ \partial L}}{ \partial
\dot{\theta} ^a}. \eqno     (2.2a) $$

We shall use left derivatives defined as follows:

$$ \frac{\overrightarrow{\partial}}{\partial \theta^a} \theta^b
f= \delta^b_a f - \theta^b
\frac{\overrightarrow{\partial}}{\partial \theta^a} f . \eqno
(2.3) $$

By defining the generalized coordinates

$$ \tilde{a} ^a := i(p^{\theta a} - i \theta^a) \;,\;
\tilde{\tilde{a}}{}^a := -(p^{\theta a} + i \theta^a). \eqno
(2.4) $$

we may write

$$ S^{a b} = \tilde{S} ^{a b} + \tilde{\tilde{S}}{}^{a b} \;,\;
\tilde{S} ^{a b} = - \frac{i}{4} ( \tilde{a} ^a \tilde{a} ^b -
\tilde{a} ^b \tilde{a} ^a) \;,\; \tilde{\tilde{S}}{}^{a b} = -
\frac{i}{4} (\tilde{\tilde{a}}{}^a \tilde{\tilde{a}}{}^b
-\tilde{\tilde{a}}{}^b \tilde{\tilde{a}}{}^a ) . \eqno      (2.5) $$

Either $L^{ab}$ or $S^{ab}$ or $\tilde{S}^{ab}$ or
$\tilde{\tilde{S}}{}^{ab}$ form the Lie algebra of the Lorentz group.
It appears$ ^{1}$ that $S^{ab}$ define vector and scalar
representations while $\tilde{S}^{ab}$ and $\tilde{\tilde{S}}{}^{ab}$
define spinor representations in the Grassmann space. The
Hamilton function $ H =
\dot{x}^{\mu} p_{\mu} + \dot{\theta}^{\mu} p^{\theta}_{\mu} - L $
defines the Hamilton equations: $\frac{\partial H }{\partial
x^{\mu}}=  - \dot{p}_{\mu} , \frac{\partial H }{\partial p_{\mu}}=
\dot{x}^{\mu} , \frac{\overrightarrow{\partial} H }{\partial
\theta^{\mu}} = - p^{\theta} { }_{\mu} ,
\frac{\overrightarrow{{\small \partial}} H}{\partial p^{\theta} {
}_{\mu}} = - \dot{\theta}^{\mu} $ and the Poisson brackets
in the ordinary and the Grassmann space :

$$\{B,A\}_p= -
\frac{ \partial A}{ \partial x^a} \frac{ \partial B}{ \partial
p_a}  + \frac{ \partial A}{ \partial p_a} \frac{ \partial B}{ \partial
x^a} - ( \frac{ \overrightarrow{ \partial A}}{\partial \theta
^a} \frac{ \overrightarrow{ \partial B}}{\partial p^\theta_a} +
  \frac{ \overrightarrow{ \partial A}}{\partial p^\theta_a}
\frac{ \overrightarrow{ \partial B}}{\partial \theta^a})
(-1)^{n_A} , \eqno (2.6) $$

where $n_{A}$ is either one or two depending on whether A has on
odd or an even Grassmann character, respectively.
It may be checked that the Poisson brackets have the following
properties:

$$\{ A,B \}_p = (-1)^{n_A n_B +1} \{ B, A \} \;,\; \{A, BC \}_p
= \{A,B \}_p C + (-1)^{n_A n_B} B \{A,C \}_p \;, $$
$$ \{AB,C \}_p = A \{B,C \}_p + (-1)^{ n_B n_C} \{A,C \}_p B
\eqno (2.6a) $$

and fulfil the Jacobi's identity:

$$ (-1)^{ n_A n_C } \{A, \{B,C \}_p \}_p + (-1)^{n_C n_B} \{C,
\{A,B \}_p \}_p + (-1)^{n_B n_A} \{B, \{C,A \}_p \}_p = 0. \eqno
     (2.6b) $$

In the quantization procedure $ - i\lbrace A, B \rbrace_{p} $ goes
to either commutators or to anticommutators, according to the
Poisson brackets (2.6).

\vspace{5mm}

2.1 SPINORS

\vspace{5mm}
Defining a particle supergeodesics by supercoordinates$^{1-2}$
$X^{a} = x^{a} + \varepsilon \xi \theta^{a}$ , which depend on
two parameters: on an ordinary time parameter $\tau$ and on a
Grassmann odd parameter $\xi$, $\varepsilon$ is here an ordinary
complex coodinate, the action follows

$$I= \frac{1}{2} \int d \tau d \xi E D_A X^a D_B X^b \eta_{a b}
\eta^{A B} , \eqno    (2.7) $$

where $D_{A} = E^{i}_{A}\partial_{i}$,
$\partial_{i} : = (\partial_{\tau}, \overrightarrow{\partial} _{\xi})$,
$\tau^{i} = (\tau , \xi)$ may be written which is invariant
under general coordinate transformations in a two dimensional
superspace $\tau^{i}$ and under the Lorentz transformations in a
2d-dimensional superspace. By choosing $\eta_{AA} = 0, \eta_{12}
= \eta_{21} = 1$ , and

$$ E^i_A = \left( \begin{array}{cc}
1 \;, & - \epsilon M\\
\xi \;, & N- \xi M
 \end{array} \right) \;,\; E=\frac{1}{N}, \eqno (2.8)$$

while $\eta_{ab} = (1,-1,-1,-1,-1,...-1)$ and integrating the
action (2.7) over the Grassmann odd parameter $\xi$ ,
the action for a superparticle follows

$$I= \int d \tau (\frac{1}{N} \dot{x} ^a \dot{x} _a + \epsilon^2
\dot{\theta} ^a \theta_a - \frac{2 \epsilon^2 M}{N} \dot{x} ^a
\theta_a) ,  \eqno    (2.7a) $$

which requires that the coordinate in
the Grassmann space is
proportional to its conjugate momentum
$$ p^\theta_a := \frac{\overrightarrow{\partial L}}{ \partial
\dot{\theta} ^a} = \epsilon^2 \theta ^a,  \eqno      (2.9) $$

bringing into the theory the spinorial degrees of freedom. For
$\varepsilon^{2}  = -i $ , it follows that
$\tilde{\tilde{a}}{}^{a} = 0$
and so are $ \tilde{\tilde{S}}{}^{ab} = 0$ , while $S^{ab} =
\tilde{S}^{ab}$.

The variation of the action (2.7a) with respect to M and N,
the former having an odd the later an even Grassmann character,
gives the two constraints:
$$ p^a \tilde{a} _a = 0 = p^a p_a .  \eqno   (2.10)$$

which , according to the Poisson brackets(2.6), in the
quantization procedure define the Dirac and the Klein-Gordon
equation, respectively. The operators $\theta^{a}, p^{\theta a}$ (in the
coordinate representation $\theta^{a} \longrightarrow \theta^{a} ,
p^{\theta}_{a} \longrightarrow -i \frac{\partial}{\partial \theta^{a}}$)
fulfil the Grassmann algebra, while the operators $\tilde{a}^{a}
$
and $\tilde{\tilde{a}}{}^{a}  $fulfil the Clifford algebra:

$$ \{\tilde{a}{ }^{a} , \tilde{a}{ }^{b}\}_{+} = 2\eta^{ab} =
\{\tilde{\tilde{a}}{ }^{a} , \tilde{\tilde{a}}{ }^{b} \}_{+} ,
\eqno (2.11) $$

with $ \{\tilde{a}{ }^{a} , \tilde{\tilde{a}} { }^{b} \} = 0 =
\{ \tilde{S}{ }^{ab} , \tilde{\tilde{S}}{ }^{cd} \}_{-} $ and
 $\tilde{S}^{ab} = - \frac{i}{4}\lbrack \tilde{a}{ }^{a} ,
\tilde{a}{ }^b \rbrack_{-} , \tilde{\tilde{S}}{ }^{ab} =-\frac{i}{4}
\lbrack \tilde{\tilde{a}}{ }^{a} , \tilde{\tilde{a}}{ }^{b}
\rbrack_{-} $. The constraints (2.10) lead to the Dirac and the
Klein-Gordon equation

$$ p^a \tilde{a} _a | \tilde{\Psi} > = 0 \;,\; p^a p_a |
\tilde{\Psi}> = 0 . \eqno  (2.12) $$

In the case that $ <\tilde{\psi}|p^{5}| \tilde{\psi} >= m $ and $ d=5$,
 it follows

$$ (\tilde{a}{ }^{b} p_{b} - \tilde{a}{ }^{5} p^5 )|\tilde{\psi} > =
0 = (\tilde{\gamma}{ }^b p_{b} - m )|\tilde{\psi}> \;,\;
b=0,1,2,3. \eqno  (2.13) $$

Since the Dirac $ \gamma^{b}$  operators have an
even Grassmann character, we assume that

$$\tilde{\gamma} ^b = - \tilde{a} ^5 \tilde{a} ^b = - 2i
\tilde{S} ^{5b} \;,\; b=0,1,2,3  \eqno  (2.14)$$

are the Dirac operators. It can be checked that in the
four-dimensional subspace $\tilde{\gamma}{ }^{b} $ fulfil the
Clifford algebra $\{\tilde{\gamma}{ }^{b} , \tilde{\gamma}{
}^{c}\}  = \eta{^{ab}} $ , while $ \tilde{S}{ }^{cd} = -\frac{i}{4}
\lbrack \tilde{\gamma}{ }^{c},\tilde{\gamma}{ }^{d}\rbrack_{-} ,
c,d = 0,1,2,3$. The existence of $\tilde{\gamma}{ }^{c}$ as an
even Grassmann operator requires that $d\ge 5$ . The operator
$\tilde{\Gamma}=i \tilde{a}{ }^{0} \tilde{a}{ }^{1} \tilde{a}{
}^{2} \tilde{a}{ }^{3} = i \tilde{\gamma}{ }^{0} \tilde{\gamma}{ }
^{1} \tilde{\gamma}{ }^{2} \tilde{\gamma}{ }^{3} $ is one of the
two Casimir operators of the Lorentz group.

There are $2^d$ products of operators $ \tilde{a}^{\mu} $ which
form the Clifford algebra, half of them with an even Grassmann
character form by themselves the Dirac algebra. Generators of the
infinitesimal translations in the ordinary space $ p^{\mu} $
and in the Grassmann space $\tilde{a}^{\mu}$ and the generators
of the Lorentz transformations in both spaces $\tilde{M}^{\mu
\nu}$ form the super-Poincar\' e algebra$^{1}$. The super-Pauli-
Ljubanski vector $ \tilde{G}^{\mu} = \frac{1}{3!}
\varepsilon^{\mu} { }_{\alpha \beta \gamma \rho }
\tilde{M}^{\alpha \beta} \tilde{a}^{\gamma} p^{\rho}$ can be
defined so that $ \tilde{G}^{\mu} \tilde{G}_{\mu} , p^{\mu} p_{\mu}$
and $ \tilde{a}^{\mu} \tilde{a}_{\mu} $ commute with all generators of
the super-Poincar\' e group as well as with $ \tilde{G}^{\rho}$
itself. The super-Pauli-Ljubanski vector define spinor
charges$^{1}$ , which being applied on a vacuum scalar state
define spinors.

\vspace{5mm}

2.2 SCALARS AND VECTORS

\vspace{5mm}

Dynamics of spinors occurs  their momenta in the Grassmann
space is proportional to their coordinates. Dynamics of scalars and
vectors occur when their momenta in the Grassmann space are
proportional to the derivative of  Grassmann coordinates with
respect to the time parameter $\dot{\theta}^{\mu} $. The
constraints then lead to the Klein-Gordon equation. In this case
the generators of the Lorentz transformations and the
translations in the Grassmann space are $ S^{\mu \nu} = {\theta}
^{\mu} p^{\theta \nu} - {\theta}^{\nu} p^{\theta \mu}$ and
$ p^{\theta \mu}$, respectively.

\vspace{5mm}

3. A PARTICLE IN A GRAVITATIONAL FIELD

\vspace{5mm}

We suppose $^{1,4}$ that supervielbeins transform vectors of a
freely falling coordinate system into vectors of an external
coordinate system. Due to two types of derivatives
$\partial_{i} \; (i=1,2)$ (eq.2.7) we assume two types of
supervielbeins: ${\bf e}^{ia} _{\mu}, i=1,2$ , the index a refers
to a freely falling coordinate system(a Lorentz index), the
index $\mu$ refers to external coordinate system(an Einstein
index). Supervielbeins depend on ordinary and Grassmann
coordinates.
We write

$$ \partial_i X^a= {\bf e}^{i a} { }_{\mu} \partial_i X^{\mu} \;,\;
\partial_i X^{\mu} = {\bf f}^{i \mu} { }_a \partial_i X^a \;,\;
i=1,2. \eqno   (3.1)$$

It follows(eq.(3.1))that

$$ {\bf e}^{i a} { }_{\mu} {\bf f}^{i \mu} { }_b = \delta^a_b \;,\;
{\bf f}^{i \mu} { }_{a} {\bf e}^{i a} { }_{\nu} = \delta^{\mu}_{\nu} .
\eqno (3.2) $$

Supervielbeins are vectors with respect to the Lorentz and the
Einstein index. Making a Taylor expansion of supervielbeins in
terms of $\xi$ $(\xi^{2}=0)$ we find

$$ {\bf e}^{i a} { }_{\mu} = e^{i a} { }_{\mu} + \varepsilon \xi \theta^b
e^{i a} { }_{ \mu b} \;,\; {\bf f}^{i \mu} { }_a = f^{i \mu} {
}_a - \varepsilon \xi \theta^b
f^{i \mu} { }_{a b} \;,\;i=1,2.  \eqno (3.3) $$

Both expansion coefficients are fields, which depend on ordinary
and Grassmann coordinates. While $ e^{ia} { }_{\mu}$ have an even
Grassmann character and describe the spin 2 part of a
gravitational field,
$ \varepsilon \theta^{b} e^{ia} { }_{\mu b}$ have an odd Grassmann
character ($\varepsilon$ is a complex constant) and are
as superpartners of $e^{ia} { }_{\mu }$ candidates for spinorial part
of a gravitational field.

{}From eqs(3.2) and (3.3) it follows that

$$   e^{i a} { }_{\mu} f^{i \mu} { }_b = \delta^a { }_b \;,\;
f^{i \mu} { }_{a}  e^{i a} { }_{\nu} = \delta^{\mu} { }_{\nu}  \;,\;
e^{i a} { }_{\mu b} f^{i \mu} { }_c = e^{i a} { }_{\mu} f^{i
\mu} { }_{c b} \;,\; i=1,2. \eqno (3.2a) $$

We find  super tensors
${\bf g}^{i}_{\mu \nu} = {\bf e}^{ia} { }_{\mu} {\bf e}^{i}_{a \nu},
{\bf g}^{i \mu \nu} ={\bf f}^{i \mu} { }_{a} {\bf f}^{i \nu a} ,
i=1,2$  with an even Grassmann character and the properties
${\bf g}^{i \mu \sigma} {\bf g}^{i}_{\sigma \nu} = \delta ^{\mu}
{ }_{\nu}= g^{i \mu \sigma} g ^{i}_{\sigma \nu}$, with
$g^{i}_{\mu \sigma} = e^{ia} { }_{\mu} e^{i} { }_{a \sigma} $.

We see from eq.(3.1) that vectors in an ordinary and a Grassmann
space are connected as follows

$$ \dot{x}^a= e^{1 a} { }_{\mu} \dot{x}^{\mu} \;,\; \dot{x}^{\mu} =
f^{1 \mu} { }_a \dot{x}^a\;,\; \theta^a=e^{2 a} { }_{\mu}
\theta^{\mu} \;,\; \theta_{\mu} = f^{2 \mu} { }_a \theta^a, \eqno  (3.4) $$

$$ \dot{\theta}^a= e^{1 a} { }_{\mu} \dot{\theta}^{\mu} + \theta^b
e^{1 a} { }_{\mu b} \dot{x}^{\mu} = (e^{2 a} { }_{\mu}
\theta^{\mu})^. =  e^{2 a} { }_{\nu , \mu_x} \dot{x}^\mu
\theta^{\nu} + e^{2 a} { }_{\mu} \dot{\theta}^{\mu} + \dot{\theta}^{\mu}
\overrightarrow{e^{2 a}} { }_{\nu ,\mu_{\theta} } \theta^{\nu}. $$

We use the notation $e^{2a} { }_{\nu,\mu_{x}} = \frac{\partial}{
\partial x{^\mu}} e{^2a}_{\nu},
\overrightarrow{e}^{2a} { }_{\nu,\mu^{\theta}} =
\frac{\overrightarrow{\partial}}{\partial \theta^{\mu}}
e^{2a} { }_{\nu}$ .

Eq.(3.4) defines the following relations among fields

$$ e^{ 2 a} { }_{\mu b}=0 \;,\;
 \overrightarrow{e^{2 a}} { }_{\nu , \mu_{\theta}} \theta^{\nu} =
e^{1 a} { }_{\mu} - e^{2 a} { }_{\mu} \;,\;
 e^{1 a} { }_{\mu b} = e^{2 a} { }_{\nu , \mu_x} f^{2 \nu} {
}_b, \eqno (3.4a)$$

which means that a point particle with a spin sees a spin
connection $ \theta^{b} e^{ia} { }_{\mu b} $ related to a vielbein
$ e^{2a} { }_{\nu}$.

Rewritting the action(2.7) in terms of an external coordinate system
according to eqs.(3.1), using the Taylor expansion of
supercoordinates $ X^{\mu}$ and superfields $ \bf{e}^{ia} { }_{\mu}$ and
integrating the action over the Grassmann odd parameter $\xi$
the action

$$ I=\int d\tau \{ \frac{1}{N} g^1_{\mu \nu} \dot{x}^\mu
\dot{x}^\nu - \epsilon^2 \frac{ 2 M}{N} \theta_a e^{1 a} { }_{\mu}
\dot{x}^\mu + \varepsilon^2 \frac{1}{2}( \dot{\theta}^\mu
\theta_a -\theta_a \dot{\theta}^\mu) e^{1 a} { }_{\mu} + \varepsilon^2
\frac{1}{2} (\theta^b \theta_a -\theta_{a} \theta^b ) e^{1
a} { }_{  \mu b} \dot{x}^\mu \} ,$$
$$\eqno  (3.5) $$

defines the two momenta of the system

$$ p_{\mu}=\frac{\partial L}{\partial \dot{x}^\mu}= p_{0 \mu} +
\varepsilon^2 \theta^a \theta^b e^1_{a \mu b} \;,\; p_{0 \mu}=
\frac{2}{N} (\dot{x}_\mu - \varepsilon^2 M p^\theta_\mu ), $$
$$ p^\theta_\mu= \varepsilon^2 \theta_a e^{1 a} { }_{\mu} = \varepsilon^2
(\theta_\mu + \overrightarrow{e^{2 a}} { }_{\nu , \mu_{\theta}}
e^2 { }_{a \alpha} \theta^{\nu} \theta^{\alpha}) . \eqno (3.6)$$

For $ p^{\theta}_{a} = p^{\theta}_{\mu} f^{1 \mu} { }_{a}$ it follows
that $ p^{\theta}_{a}$ is proportional to $\theta_{a}$. For a
choice $\varepsilon^{2} = - i $, $ \tilde{a}_{a} = i
(p^{\theta}_{a} - i \theta_{a}), \tilde{a}^{\mu} = f^{1\mu} { }_{a},
 $ while $ \tilde{\tilde{a}}_{a}= 0 $. In this case me may write

$$ p_\mu= p_{0 \mu} + \frac{1}{2} \tilde{S}^{a b} e^1_{a \mu b}
= p_{0 \mu} + \frac{1}{2} \tilde{S}^{a b} \omega_{a b \mu} \;,\;
\omega_{a b \mu}=\frac{1}{2} (e^1_{a \mu b} - e^1_{b \mu a}).
\eqno (3.6a) $$

 and obtain the Hamilton function

$$ H=\frac{N}{4} g^{1 \mu \nu} p_{0 \mu} p_{0 \nu} + \frac{i}{2}
M p_{0 \mu} f^{1 \mu} { }_a \tilde{a}^a . \eqno   (3.7)$$

and the two constraints

$$ p_0^\mu p_{0 \mu} = 0 = p_{0 \mu} f^{1 \mu} { }_a \tilde{a}^a .
\eqno (3.8)$$

In the quantization procedure the two constraints in eqs.(3.8)
$ p_{0\mu} f^{1\mu} { }_{a} \tilde{\gamma}^{a} p_{0\nu} f^{1\nu}
 { }_{b} \tilde{\gamma}^{b} = 0 = p_{0\mu} f^{1\mu} { }_{a} \tilde
{a}^{a}$ have to be symmetrized properly, due to the fact that
fields depend on ordinary$^{5}$ and Grassmann coordinates,
in order that the Klein-Gordon and the Dirac equations in the
presence of gravitational fields follow correspondingly.

A torsion and a curvature follow from the Poisson brackets
$ \{ p_{0a},  p_{0b} \}$, with $ p_{0a} = f^{1\mu} { }_{a} ( p_{\mu} +
\frac{1}{2} \tilde{S}^{cd} \omega_{cd\mu}) $.

We find

$$ \{ p_{0 a} , p_{0 b} \}_p = -\frac{1}{2} S^{c d} R_{ c d a b}
+ p_{0 c} T^c_{a b} , \eqno (3.9)$$
$$ R_{c d a b} = f^{1 \mu} {}_{[a} f^{1 \nu} {}_{b]}
( \omega_{cd\nu,\mu^{x} } + \omega_{c} {}^e{}_{ \mu} \omega_{e d \nu}
\overrightarrow{\omega}_{c d \mu , f^{\theta}} \theta^e \omega_e
{ }^f_{ \nu}), $$
$$ T^c { }_{a b} =e^1 {}_{c \mu} ( f^{1 \nu} {}_{[b} f^{1 \mu} {}_{ a]} {}_{,
\nu} + \omega_{e \nu} { }^d \theta^e f^{1 \nu} {}_{[b}
\overrightarrow{f^{1 \mu}} {}_{  a]} {}_{, d^\theta} ),$$
$$ {\rm with } \; A _{[a} B _{b]} = A_a B_b - A_b B_a. $$

If the action for a free gravitational field is

$$ I=\int d^{d} x d^{d}\theta \omega {\cal L}, \eqno(3.10) $$

where $\omega$ is a scalar density in the Grassmann space$^{1}$,
the Lagrange density $ {\cal L}$
includes  $ det(e^{1a} { }_{\mu})  R $, $ R =  R^{ab}
 {}_{ab} $, or(and) $ det(e^{1a} { }_{\mu})  T ^a { }_{cd}
 T^{cd} { }_a$.

Generators of the Lorentz transformations in the Grassmann space
form a group $ SO(1,N-1)$ which can be decomposed$^6$ ($ SO(1,17) $
for example) into $ SO(1,3)\times SU(2)\times SU(3)\times...$.
This means that the generators of the Lorentz transformations in
the Grassmann space in higher then four dimensions may define
weak and colour charges, while vielbeins and spin connections
with the Lorentz index higher then four manifest in the
four-dimensional ordinary subspace as weak and colour fields$^3$.
Since the momentum in the fifth dimension of the ordinary space
manifests as an electromagnetic charge of a particle in the
four-dimensional ordinary subspace and accordingly the spin
connection with the Lorentz index equal to five manifests as an
electromagnetic field, can Lorentz indices only greater then five
be connected with fields, which manifest in the four-dimensional
subspace as non-Abealen  gauge fields.

\vspace{5mm}

4.ELECTRODYNAMICS AS A GRAVITATION IN THE FIFTH DIMENSION

\vspace{5mm}

We shall present in this section how a vielbein and a spin
connection, depending on the ordinary and Grassmann coordinates,
with the Lorentz index equal to five manifest in the
four-dimensional subspace as an electromagnetic field. The
dependence of the field on the Grassmann coordinates determines
the internal spin one of the field $^{1}$.

We shall treat the case with $ d=5 $ and with no gravitational field
in the ordinary four-dimensional subspace:

$$e^{ i m} { }_{\mu} = \delta^m { }_{\mu} \;,\; e^{i 5} {
}_{\alpha} = e^{i
5} { }_{\alpha} (x^\beta, \theta^\mu) \;,\; e^{i 5} { }_5 =
e^{i 5} { }_5 (
x^\beta, \theta^\mu), \eqno (4.1a)$$
$$ m=0,1,2,3 \;;\; \alpha, \beta, \gamma = 0,1,2,3 \;;\;
\mu=\alpha,5 \;;\; i=1,2 .$$

It follows then from eq.(3.2) that

$$f^{i \alpha} { }_b = \delta^\alpha { }_\beta \;,\; f^{i 5} { }_m =
-(e^5 { }_5)^{-1} e^{i 5} { }_m \;,\; f^{i 5} { }_5 = (e^5 {
}_5)^{-1} . \eqno  (4.1b)$$

If we take into account the choice of the fields of eqs.(4.1)
in the action(3.5) and in eq.(3.6a) we find a covariant momenta

$$ p_{0 \alpha} = p_\alpha - e^{1 5} { }_\alpha e^{1 5} { }_5
{}^{-1}  p_5 - \frac{i}{4} \tilde{\gamma}^a e^2 { }_{ 5[a,
\alpha^x]} \;,\; \alpha=0,1,2,3 , \eqno (4.2)$$

with $p_{0 \mu} = \frac{2}{N} e^{15} { }_{5} ( e^{1} { }_{5\nu}
\dot{x}^{\nu} + \frac{i}{2} M e^{1a} { }_{5} \tilde{a}_{a}) $ taken
as a constant. Indices a as well as $\alpha$ are, due to the
supposition (4.1a), raised or lowered by the Minkowski matrix
$\eta_{\alpha , \beta}$.

The Hamilton function

$$H = \frac{N}{4} p_{0 \alpha} p^{0 \alpha} + \frac{i}{2} M
p_0^\alpha \tilde{a}_\alpha \;,\; \alpha=0,1,2,3 \eqno (4.3)$$

and the two constraints can be found

$$ p_0^\alpha \tilde{a}_\alpha - m \tilde{a}^5 = 0 =
p_{0}^{\alpha} p_{0 \alpha} - m^2 = 0 , \eqno (4.4a,b)$$

where $ m= - p_{5} e^{15} { }_{5} {}^{-1} $.

The two fields $ e^{i5} { }_{\alpha}, i=1,2 $, depending on ordinary
and Grassmann coordinates, have an even Grassmann character and
are Lorentz vectors. They are related to each other and to the
field of an odd Grassmann character $ \theta^{b} e^{1a} { }_{\alpha
b}$ (3.4a)

$$ e^{1 5} { }_{\alpha} = e^{ 2 5} { }_{\alpha} + \overrightarrow{e^{2
5}} { }_{\nu, \beta^{\theta}} \theta^{\nu}$$
$$e^{1 5} { }_{\alpha \beta} = e^{2 5} { }_{\beta, \alpha^x} + e^{2
5} { }_{5, \alpha^x} f^{2 5} { }_{\beta}\;,\; e^{1 5} { }_{5 \alpha}=0.
\eqno (4.5)$$

In the gauge $^{1}$  in which

$$ e^{-\frac{i}{2} \omega_{\alpha \beta} S^{\alpha \beta}}
e^{25} { }_ {\alpha} e^{\frac{i}{2} \omega_{\alpha \beta}
S^{\alpha \beta}} = \Lambda_{\alpha} { }^{\beta}
 e^{25} { }_{\beta} ,
 \eqno (4.5a)$$

with $ S^{\mu \nu}$ being the differential operator
in the Grassmann space defined in eq.(2.2) (after quantization)
and $\Lambda^{\alpha} { }_{\beta}$ a tensor of the Lorentz
transformations , it follows$^{1}$ that $ e^ {25} { }_{\alpha} =
\frac{1}{2} e^{15} { }_{\alpha}$.
We may write

$$e^{1 5} { }_{\alpha} e^{1 5} { }_5 {}^{-1} p_5 = e {\cal A}_{\alpha} .
\eqno (4.6) $$

The first constraint of eqs.(4.4a) can then be written in the
form

$$ \frac{1}{2} [ \tilde{\gamma}^\alpha(p_{\alpha} + e {\cal A}_
\alpha ) - \frac{e}{4 m} \tilde{S}^{\alpha \beta} {\cal
F}_{\alpha \beta} + (p_{\alpha} + e {\cal A}_\alpha )
\tilde{\gamma}^{\alpha} - \frac{e}{4 m} {\cal F}_{\alpha \beta}
\tilde{S}^{\alpha \beta} ] - m = 0 . \eqno (4.7)$$

In the quantization procedure(according to the Poisson
brackets(2.6) ), eq.(4.7) has to be symmetrized
since the field depends on ordinary and Grassmann coordinates.

In a nonhomogenous field the term with $ {\cal F}_{\alpha
\beta}$ causes an anomalous magnetic moment of a charged
particle, unless the particle is considered as a (almost
massless) cluster of very heavy
constituents $^{3}$. In an homogenous field the contribution of
that term vanishes since $\lbrack p_{\alpha}, e {\cal A}_{\alpha} +
\frac{i}{8} \tilde{\gamma}^{\beta} {\cal F}_{\alpha \beta} \rbrack =
e {\cal A}_{\beta,\alpha} $.

The Poisson brackets of eq.(3.9) bring
in the gauge (4.5a) a torsion

$$ T^m { }_{a b} = 0 \;,\; T^5 { }_{m n} = {\cal F}_{m
n} - \omega_m {}^5 {}_n e^5 {}_5
{}^{-1} \eqno   (4.8)$$

and a Riemann tensor which depends on $\omega_{mn\mu} $ only.

The variation of the action (3.10) with $ {\cal L} =  T^{5}
{ }_{mn} T^{mn} { }_5 $ with respect to ${\cal A}_{\alpha}
$  and $\omega_{m} { }^{5} { }_{n}$
brings the ordinary Maxwell equations for a free field and the
constraints (4.5) in the gauge(4.5a). This is the solution also
for the case that $ {\cal L}=  R^{mn} { }_{mn}$.

\vspace{5mm}

5.CONCLUSIONS

\vspace{5mm}

The theory in which the space has d ordinary and d
Grassmann coordinates possesses the supersymmetry and
enables not only the canonical quantization of coordinates and
fields but offers also the possibility of unifying gauge fields:
generators of the super-Poincar\' e algebra (in the ordinary and
the Grassmann space) define
internal  degrees of freedom: spins of particles and fields and
charges. Spins of particles and fields appear as the
dynamics of particles and fields in the Grassmann
five-dimensional subspace. Momentum in the fifth ordinary
dimension manifests as a charge(and a mass) of particles and
fields in a four dimensional ordinary subspace. Generators of
the Lorentz transformations in higher then five dimensions
manifest in the four dimensional subspace charges of non-Abelian gauge
fields, while gravitational fields in higher dimensions manifest in four
dimensional subspace as corresponding fields. Spin connections
of an odd Grassmann character apear as superpartners of
vielbeins with an even Grassmann chracter. Both depend on
ordinary and Grassmann coordinates, the later defining the spin
of fields.
Generators of translations and the Lorentz transformations in
the ordinary and the Grassmann space define super- Poincar\' e
algebra, while super- Pauli- Ljubanski vector defines spinor
charges. For d=5 there are four four-spinors defined by
super-Pauli-Ljubanski vector and by generators of the Lorentz
transformations in the Grassmann space. Half of them have a
negative and half of them a positive internal parity , defined by
discrete Lorentz transformations in the Grassmann space. If
these degrees of freedom manifest on the level of quarks and
leptons, the two generations of quarks and leptons would have a
positive and the two a negative internal parity. There are also
two scalars, two three vectors and two four vectors, which
define ineternal spins of gauge fields.
\vspace{5mm}

8.APPENDIX - SUPER POINCAR\' e ALGEBRA

\vspace{5mm}

The operators $ \tilde{M}^{\mu \nu}= L^{\mu \nu}+ \tilde{S}^{\mu
\nu} $ ,$ p^\mu $,$ \tilde{a}^\mu$
from the super-Poincar\' e algebra $^1$.

$$\{ p^\mu, p^\nu \} =0 = \{ p^\mu, \tilde{a}^\nu \} \eqno
(A.1a)$$

$$\{ \tilde{M}^{\mu \nu}, p^\alpha \} = i( p^\nu \eta^{\mu
\alpha} - p^{\mu} \eta^{\nu \alpha} ) \eqno (A.1b)$$

$$\{ \tilde{a}^\mu, \tilde{a}^\nu \} = 2 \eta^{\mu \nu} \eqno
(A.1c)$$

$$\{ \tilde{M}^{\mu \nu}, \tilde{a}^\alpha \} = i( \tilde{a}^\nu \eta^{\mu
\alpha} - \tilde{a}^{\mu} \eta^{\nu \alpha} ) \eqno (A.1d)$$

$$\{ \tilde{M}^{\mu \nu}, \tilde{M}^{\alpha \beta} \} = -i(
\tilde{M}^{\mu \beta} \eta^{\nu \alpha} - \tilde{M}^{\nu \beta}
\eta^{\mu \alpha} - \tilde{M}^{\mu \alpha} \eta^{\nu \beta} +
\tilde{M}^{\nu \alpha} \eta^{\mu \beta} ). \eqno(A.1f)$$

The two Lorentz scalars can be defined : $ \tilde{I}_e =
\frac{i}{4 !} \varepsilon_{\mu \nu \rho \sigma \tau } \tilde{M}^{\mu
\nu} \tilde{M}^{\rho \sigma} p^\tau $ and $ \tilde{I}_o = -
\frac{i}{4 !} \varepsilon_{\mu \nu \rho \sigma \tau } \tilde{M}^{\mu
\nu} \tilde{M}^{\rho \sigma} \tilde{a}^\tau $, the first with an
even, the second with an odd Grassmann character. The commutator
of $ I_e$ with $ \tilde{a}^\mu $ or of $ I_o$ with $p_\mu$
defines the super-Pauli-Ljubanski vector :

$$\tilde{G}^\mu = \{ \tilde{I}_e , \tilde{a}^\mu \} = \{
\tilde{I}_o , p^\mu \} = \frac{1}{3 !} \varepsilon^\mu { }_{\alpha \beta
\gamma \rho} \tilde{M}^{\alpha \beta} \tilde{a}^{\gamma} p^\rho
= \frac{1}{3!}  \varepsilon^\mu { }_{\alpha \beta
\gamma \rho} \tilde{S}^{\alpha \beta} \tilde{a}^{\gamma} p^\rho
. \eqno(A.2) $$

It has an odd Grassmann character. We can find it as one of
constants of motion for the action(2.7a). The following properties
can be prooved:

$$\{\tilde{G}^\mu ,p^\nu\}= 0,$$
$$\{\tilde{a}^\mu ,\tilde{G}^\nu\}= - \varepsilon^{\mu \nu}
{ }_{\alpha \beta \rho} \tilde{S}^{\alpha \beta} p^\rho, $$
$$\{\tilde{M}^{\mu \nu} , \tilde{G}^\rho\}= i(\eta^{\mu \rho}
\tilde{G}^\nu - \eta^{\nu \rho} \tilde{G}^\mu), $$
$$\{\tilde{G}^\mu ,\tilde{G}^\nu\}= \frac{1}{3 !} \varepsilon^{\rho
\mu} { }_{\alpha \beta \gamma} \varepsilon_\rho { }^\nu {
}_{\acute{\alpha} \acute{\beta} \acute{\gamma}}
\tilde{S}^{\alpha \beta} p^\gamma \tilde{S}^{\acute{\alpha}
\acute{\beta}} p^{\acute{\gamma}},$$
$$\tilde{G}^\mu p_\mu = 0, \{\tilde{G}^\mu, \tilde{a}^\nu p_\nu \}=
0. \eqno(A.4) $$

There are $\tilde{G}^\mu \tilde{G}_\mu, p^\mu p_\mu $ and
$\tilde{a}^\mu \tilde{a}_\mu$ which commute with all generators of
the super-Poincar\' e group as well as with $\tilde{G}^\rho$
itself.

We may write $ \tilde{G}^\mu$ for $d=5$ in an explicit form as follows:

$$ \tilde{G}^{0} = \tilde{S}_3 \tilde{a}^3 p^{5} -
\overrightarrow{\tilde{S}} \cdot { } \overrightarrow{p}
\tilde{a}^5 , \; \overrightarrow{\tilde{G}}
=\overrightarrow{\tilde{S}}(\tilde{a}^0 p^5 - \tilde{a}^5 p^{0})
+ (\overrightarrow{\tilde{K}}
\times \overrightarrow{p}) \tilde{a}^5,$$

$$\tilde{G}^5 = \tilde{S}_3 \tilde{a}^3 p^0-
\overrightarrow{\tilde{S}} \cdot { } \overrightarrow{p} \tilde{a}^0, \;
{ \rm with} \; \tilde{S}_i =\frac{1}{2} \varepsilon_{i j k} \tilde{S}
^{jk}, \tilde{K}_i = \tilde{S}^{0i}. \eqno(A.5)$$

Being applied on a vacuum state which is a scalar in the
Grassmann space, $\tilde{G}^\mu$ produces spinors. For the case
that a scalar fulfils the Klein-Gordon equation and in  the
representation in which $ p^\mu = (p^0 ,0,0,p^3,0) $ the two
operators $$\tilde{Q_{1,2}}=\frac{\mp 1}{\sqrt{|p_0|}}
(\tilde{G}_1 \pm i \tilde{G}_2), \eqno(A.6) $$

define four spinors according to the choice of either $ p^0 =
p^3$ or $ p^0= - p^3$ . We define accordingly four
operators $ \tilde{Q}^{\pm}{ }_{1,2} $. The four spinors, which are
chosen to be eigensatates of $\tilde{S}_3 $
and fulfil the Dirac equation $p^\mu \tilde{a}_\mu
 \tilde{Q}^{\pm} { }_{1,2}=0 $ are presented on Table I.

All vectors of an even Grassmann character can be obtained from
spinors by the application of an operator of an odd Grassmann
character. On Table II the eigenvectors of the two Casimir
opertors $\frac{1}{2}\tilde{S}^{\alpha \beta} \tilde{S}_{\alpha \beta}$ and
$\Gamma = \frac{-i}{3!} \varepsilon^{\alpha \beta \gamma \delta}
S_{\alpha \beta} S_{\gamma \delta}, \alpha,\beta \cdots =
0,1,2,3 $ and $ S_3 $ are presented which are for $d=5$ two
scalars, two three-vectors and two four-vectors.\\

Table I : The four spinors generated by the operators
$\tilde{Q}^{\pm} { }_{1,2} $ on a scalar vacuum state are
presented(printed in bold face) together with the rest of
spinors, which form the Weyl four spinors .

\vspace{5mm}

\begin{center}
\begin{tabular}{|l|l|} \hline
${\bf \bf<\theta | ^1 \varphi1 > = \frac{1}{2}(\theta^1 + i \theta^2)
(\theta^0 - \theta^3) \theta^5}$ & $<\theta | ^5 \varphi1 > =
\frac{1}{2}(1 + i \theta^1 \theta^2) (\theta^0 - \theta^3) $ \\
\hline
$<\theta | ^1 \varphi2 > = -\frac{1}{2}(1-i \theta^1 \theta^2)
(1- \theta^0  \theta^3) \theta^5$ & $<\theta | ^5 \varphi2 > =
- \frac{1}{2}( \theta^1 - i \theta^2) (1- \theta^0 \theta^3) $ \\
\hline
$<\theta | ^2 \varphi1 > = -\frac{1}{2}(\theta^1 + i \theta^2)
(1- \theta^0  \theta^3) $ & $<\theta | ^6 \varphi1 > =
\frac{1}{2}(1 + i \theta^1 \theta^2) (1-\theta^0 \theta^3) \theta^5$ \\
\hline
$<\theta | ^2 \varphi2 > = -\frac{1}{2}(1-i\theta^1  \theta^2)
(\theta^0 - \theta^3) $ & ${\bf<\theta | ^6 \varphi2 > =
\frac{1}{2}( \theta^1 - i \theta^2) (\theta^0 - \theta^3) \theta^5}$ \\
\hline \hline
$<\theta | ^3 \varphi1 > = \frac{1}{2}(\theta^1 + i \theta^2)
(1+\theta^0 \theta^3) $ & $<\theta | ^7 \varphi1 > =
\frac{1}{2}(1 + i \theta^1 \theta^2) (1+ i \theta^0  \theta^3) \theta^5$ \\
\hline
$<\theta | ^3 \varphi2 > = -\frac{1}{2}(1-i\theta^1  \theta^2)
(\theta^0 + \theta^3) $ & ${\bf<\theta | ^7 \varphi2 > =
-\frac{1}{2}( \theta^1 - i \theta^2) (\theta^0 + \theta^3) \theta^5} $ \\
\hline
${\bf<\theta | ^4 \varphi1 > = -\frac{1}{2}(\theta^1 + i \theta^2)
(\theta^0 + \theta^3) \theta^5}$ & $<\theta | ^8 \varphi1 > =
\frac{1}{2}(1 + i \theta^1 \theta^2) (\theta^0 + \theta^3) $ \\
\hline
$<\theta | ^4 \varphi2 > = -\frac{1}{2}(1- i \theta^1 \theta^2)
(1+\theta^0  \theta^3) \theta^5$ & $<\theta | ^8 \varphi2 > =
\frac{1}{2}(\theta^1 - i \theta^2) (1+\theta^0  \theta^3) $ \\
\hline
\end{tabular}
\end{center}

\vspace{5mm}

Table II : The two scalars, two three-vectors and two
four-vectors for d=5 are presented. They are defined by the
Casimir operators for the generators of the Lorentz
transformations of the vectorial character in the Grassmann space.

\vspace{5mm}

\begin{center}
\begin{tabular}{|l|l|} \hline
$<\theta | ^1 \varphi1 > = 1 $ & $<\theta | ^5 \varphi1 > =
\frac{1}{2}(\theta^1 + i \theta^2) (1 + \theta^0 \theta^3)
\theta^5$\\ \hline
$<\theta | ^1 \varphi1 > = \theta^0 \theta^1 \theta^2 \theta^3 $
& $<\theta | ^5 \varphi2 > =
\frac{1}{2}(\theta^0 +  \theta^3) (1 + i \theta^1 \theta^2) \theta^5
$\\ \hline
$<\theta | ^3 \varphi1 > =\frac{1}{2}( \theta^0- \theta^3 )(
\theta^1  + i \theta^2)$ & $<\theta | ^5 \varphi3 > =
\frac{1}{2}(1-\theta^0 \theta^3) (\theta^1 - i \theta^2) \theta^5$ \\
\hline
$<\theta | ^3 \varphi2 > = -\frac{1}{\sqrt{2}}(\theta^0 \theta^3
+ i \theta^1 \theta^2)$ & $<\theta | ^5 \varphi4 > =
\frac{1}{2}(\theta^0 +  \theta^3) (1-i\theta^1  \theta^2) \theta^5 $ \\
\hline
$<\theta | ^3 \varphi3 > =-\frac{1}{2}( \theta^0+ \theta^3 )(
\theta^1  - i \theta^2)$ & $<\theta | ^6 \varphi1 > =
\frac{1}{2}(1-\theta^0 \theta^3) (\theta^1 + i \theta^2) \theta^5$ \\
\hline
$<\theta | ^4 \varphi1 > =\frac{1}{2}( \theta^0+ \theta^3 )(
\theta^1  + i \theta^2)$ & $<\theta | ^6 \varphi2 > =
\frac{1}{2}(\theta^0 -  \theta^3) (1- i \theta^1  \theta^2) \theta^5$ \\
\hline
$<\theta | ^4 \varphi2 > = \frac{1}{\sqrt{2}}( i \theta^0 \theta^2
- \theta^0 \theta^3)$ & $<\theta | ^6 \varphi3 > =
\frac{1}{2}(1 + \theta^0 \theta^3) (\theta^1 - i  \theta^2) \theta^5 $ \\
\hline
$<\theta | ^4 \varphi3 > =-\frac{1}{2}( \theta^0- \theta^3 )(
\theta^1  - i \theta^2)$ & $<\theta | ^6 \varphi4 > =
\frac{1}{2}(\theta^0 +  \theta^3) (1+ i \theta^1  \theta^2) \theta^5$ \\
\hline
\end{tabular}\end{center}
\vspace{5mm}

7.ACKNOWLEDGEMENTS

\vspace{5mm}

The work was supported by Ministry of Science and Technology of
Slovenia and the National Science Foundation through funds
available to the US-Slovenia Joint Board for Scientific and
Technological Cooperation (no.NSF 899).

\vspace{5mm}

8.REFERENCES

\vspace{0.5cm}

\begin{description}
\item[1] N. Manko\v c -Bor\v stnik, Phys.Lett. {\bf B 292}
 (1992) 25,  Il nuovo Cimento {\bf  A 105} (1992) 1461,
 Journ. of Math. Phys. {\bf 34} (1993) 8, Int. Journal of
 Modern Phys. {\bf A 9} ( 1994) 1731,
 IC/91/369, IC/91/370, IJS.TP.92/5,
 IJS.TP.92/22, IJS.TP.93/5, IJS.TP.93/14,
 Proceedings to the Edirne conference Frontier in Theoretical
 Physics,Edirne, Dec.1993, IJS.TP.94/7

\item[2] H. Ikemori, Phys.Lett. {\bf B 199} (1987)239
\item[3] Work in progress
\item[4] J.Wess, B.Zumino, Nucl.Phys. {\bf B70} (1974) 139,
  J.Wess, J.Bagger, Supersymmetry and supergravity, Princeton Series
  in Physics (Princeton U.P., Princeton, N.J., 1983)
\item[5] B.S.De Witt, Rev. Mod.Phys.{\bf 29} (1957) 377
\item[6] J. Maalampi, M. Ross, Phys. Rep. {\bf 2} (1990) 53

\end{description}

\end{large}
\end{document}